\documentclass[a4paper,11pt,reqno]{article}

\usepackage{a4wide}
\usepackage{amsmath,amsfonts,amssymb}
\usepackage[english]{babel}
\usepackage{soul}
\usepackage{nicefrac}
\usepackage[mathscr]{euscript}
\usepackage{setspace}
\usepackage{datetime}
\usepackage[sort&compress,merge,numbers]{natbib}

\usepackage{mathrsfs}
\usepackage[T1]{fontenc}
\usepackage{mathpazo}

\usepackage[breaklinks=true]{hyperref}

\numberwithin{equation}{section}

\hypersetup{
			colorlinks=true,
			urlcolor=blue,
			citecolor=magenta,
			linkcolor=blue,
     }

\let\oldsqrt\sqrt
\def\sqrt{\mathpalette\DHLhksqrt}
\def\DHLhksqrt#1#2{%
\setbox0=\hbox{$#1\oldsqrt{#2\,}$}\dimen0=\ht0
\advance\dimen0-0.2\ht0
\setbox2=\hbox{\vrule height\ht0 depth -\dimen0}%
{\box0\lower0.4pt\box2}}

\author{
  \begin{minipage}{.97\linewidth}
    \vspace{1cm}
       \begin{center}
      \begin{small}
     \textbf{P. Marios Petropoulos}$^1$ and 
      \textbf{Konstantinos Siampos}$^2$
              \end{small}
    \end{center}
    \vspace{0.5cm}
    \hspace{2.4cm}  \begin{minipage}{.7\linewidth}
\begin{center}     {\it \begin{footnotesize}
\hbox{\kern-2.4cm\vbox{\vskip0cm
 \begin{itemize}
               \item[$^1$] Centre de Physique Th\'eorique,\\ 
        Ecole Polytechnique, CNRS,\footnote{UMR 7644}\\ 
        Universit\'e Paris-Saclay,\\
        91128 Palaiseau Cedex, France\\
        \texttt{marios@cpht.polytechnique.fr}
\vskip0.29cm
      \end{itemize}}
\kern-4cm\vbox{
\begin{itemize}
 \item[$^2$] Albert Einstein Center for Fundamental Physics,\\
Institute for Theoretical Physics,\\ 
Bern University,\\
Sidlerstrasse 5, 3012 Bern, Switzerland
\\        \texttt{siampos@itp.unibe.ch}
      \end{itemize}\vskip0.05cm
}}
     \end{footnotesize}}
\end{center}
     \end{minipage}
    \vspace{0.5cm}
  \end{minipage}
}

\title{\vspace{3.5cm}
 \boldmath 
    \textbf{Integrability, Einstein spaces and holographic fluids}
   \unboldmath
}

\begin{document}

\begin{titlepage}
\maketitle
\thispagestyle{empty}

 \vspace{-14.5cm}
  \begin{flushright}
  CPHT-RR047.0914
    \end{flushright}
 \vspace{12cm}

\begin{center}
\textsc{Abstract}\\  
\vspace{1cm}	
\begin{minipage}{1.0\linewidth}

Using holographic-fluid techniques, we discuss some aspects of the integrability properties of Einstein's equations in asymptotically anti-de Sitter spacetimes. We review and we amend the results of \href{http://arxiv.org/abs/1506.04813}{1506.04813} on how exact four-dimensional Einstein spacetimes, which are algebraically special with respect to Petrov's classification, can be reconstructed from boundary data: this is possible if the boundary metric supports a traceless, symmetric and conserved complex 
rank-two tensor, which is related to the boundary Cotton and energy--momentum tensors, and if the 
hydrodynamic congruence is shearless. We illustrate the method when the hydrodynamic congruence has vorticity and 
the boundary metric has two commuting isometries. This leads to the complete  Pleba\'nski--Demia\'nski family. 
The structure of the boundary consistency conditions depict  a $U(1)$ invariance for the boundary data, which 
is reminiscent of a Geroch-like solution-generating pattern for the bulk.

\end{minipage}
\end{center}


\end{titlepage}

\onehalfspace

\noindent\rule{\textwidth}{1.2pt}
\vspace{-1cm}
\begingroup
\hypersetup{linkcolor=black}
\tableofcontents
\endgroup
\noindent\rule{\textwidth}{1.2pt}


\section*{Introduction}
\addcontentsline{toc}{section}{Introduction}

Holographic correspondence was originally formulated at the microscopic level between type IIB string theory  on $\text{AdS}_5\times S^5$ and ${\cal N}=4$ Yang--Mills on the four-dimensional conformal boundary of 
$\text{AdS}_5$. Later on, it was extended macroscopically as a relationship between gravity plus matter in asymptotically (locally) anti-de Sitter spaces and some phenomenological boundary conformal field theory in one dimension less. The latter is usually a macroscopic quantum state, which might -- but needs not to -- be 
in the hydrodynamic regime. This is how fluids emerge in holography.

A duality correspondence is a tool for studying either of the sides of the correspondence, using knowledge and understanding available for the other. Holography has been developed in priority for unravelling the boundary quantum field theory and many attempts were made for enlarging the validity of the correspondence to quantum chromodynamics or to systems of condensed matter. On the supergravity side, the efforts were mostly focused on string theory in its non-perturbative regime, and much less for the supergravity approximation. Still, there are interesting issues to be addressed, such as the integrability properties of the gravity sector. The purpose of this note is to report on recent progress made in using holographic fluids for understanding some integrable corners of Einstein's equations.

Generically Einstein's equations are not integrable. Nevertheless, under some assumptions, the system exhibits interesting integrability properties or possesses solution-generating techniques. One of these is Geroch's \cite{Geroch, Geroch:1972yt},  generalizing a previous work by Ehlers \cite{Ehlers}. The starting point in this approach is  a four-dimensional manifold $\mathcal{M}$, endowed with a Ricci-flat metric $\mathrm{g}$ invariant under a one-parameter group of motions generated by a Killing vector $\xi$. A coset space 
$\mathcal{S}$ can be constructed as the quotient of $\mathcal{M}$ by the group of motions. The information carried by $(\mathcal{M}, \text{g}, \xi)$ is bijectively mapped onto $(\mathcal{S}, \text{h},  \text{A}, \phi)$, where  $\mathrm{h}$ is the metric 
on the projected space $\mathcal{S}$ orthogonal to $\xi$, and $(\text{A},\phi)$ the Kaluza--Klein vector and scalar fields created along with  $\mathrm{h}$. Einstein's four-dimensional dynamics for $\mathrm{g}$  translates into three-dimensional sigma-model dynamics for the fields   $(\text{h},  \text{A}, \phi)$. This sigma-model is not integrable but it exhibits a continuous group of symmetries, the $U$-duality group, which is $SL(2,\mathbb{R})$, and allows to map any solution onto another.
Although this mapping is local at the level of the three-dimensional data, no local relationship exists amongst the uplifted four-dimensional solutions. 

Following Ernst \cite{Ernst:1967wx, Ernst:1968}, Geroch's method generalizes when 2 commuting Killing vectors are available on $(\mathcal{M}, \text{g})$ and allows to reduce Einstein's dynamics to a two-dimensional sigma-model. The latter possesses full affine symmetry and is integrable \cite{Breitenlohner:1986um, belinskii, Maison:1978es, maison2, Mazur:1982}. Further generalizations of the method have been studied in great detail within supergravity theory, in various dimensions. 

As an example, according to the above pattern, the vacuum Schwarzschild Taub--NUT solution with mass $m$ and nut charge $n$, can be obtained from the pure Schwarzschild solution with mass only. More generally, the complex parameter $m+in$ is mapped onto $\text{e}^{-i\lambda}(m+in)$, under a $U(1)\subset SL(2,\mathbb{R})$ symmetry of the three-dimensional sigma-model. 

The techniques discussed here have been extensively developed over the years for vacuum or electrovacuum solutions. In the presence of a cosmological constant, Geroch's approach is more intricate though, and this case has been lesser investigated. Some recent examples
in the framework of Ernst's equations can be found in \cite{Charmousis:2006fx,Caldarelli:2008pz,Astorino:2012zm}, whereas a generalization of the plain Geroch's procedure for Einstein spaces was made available in \cite{Leigh:2014dja}. In that case the $U$-duality group turns out to be a subgroup of $SL(2,\mathbb{R})$, and the benefit for generating new solutions is limited. In particular, this group does not include the generator which realizes the mapping  $m+in \to \text{e}^{-i\lambda}(m+in)$. This is unfortunate because a full  Schwarzschild Taub--NUT solution does exist on AdS, and it is legitimate to ask whether this is indeed a manifestation of a hidden symmetry, as for the vacuum case.

The question we would like to address here is the following: \emph{does holography provide an alternative solution-generating technique?} 

Schematically, a holography-based solution-generating  technique could work as follows:
\begin{equation}
\nonumber
(\mathcal{M}, \text{g}_{\text{bulk}})  \underset{r\to \infty}{\longrightarrow}  (\mathcal{B}, \text{g}_{\text{bry.}}, \text{T}) \underset{\text{``}U\text{''}}{\rightarrow}(\mathcal{B}, \text{g}_{\text{bry.}}', \text{T}') \underset{\text{exact reconstruction}}{\longrightarrow}  (\mathcal{M}, \text{g}_{\text{bulk}}').
\end{equation}
In this pattern, the starting point is an Einstein space $(\mathcal{M}, \text{g}_{\text{bulk}})$, delivering  a boundary space $\mathcal{B}$ with boundary metric $\text{g}_{\text{bry.}}$ and boundary energy--momentum tensor $\text{T}$ ($r$ is the holographic radial coordinate).  Indeed $\text{g}_{\text{bry.}}$  and $\text{T}$ are the two holographic pieces of data for pure-gravity bulk dynamics. We can imagine that the boundary data of exact bulk Einstein spaces obey certain integrability requirements, that leave some freedom to map $(\text{g}_{\text{bry.}}, \text{T})$ into $(\text{g}_{\text{bry.}}', \text{T}')$. Reconstructing  $(\mathcal{M}, \text{g}_{\text{bulk}}')$ from $(\text{g}_{\text{bry.}}', \text{T}')$ would then lead to a possibly new exact Einstein space. 

In principle, given any two independent pieces of boundary data, one can reconstruct the bulk order by order using e.g. the Fefferman--Graham series  expansion \cite{PMP-FG1, PMP-FG2}. For arbitrary boundary data, this series expansion is generically not resummable and the bulk solution is not exact. The first task in the aforementioned program is therefore to set up the integrability requirements. Finding the group of transformations, which act on the boundary data without altering the integrability properties of the latter is the second step. 

The aim of this paper is to report on our understanding about these two steps. We will first (Sec. \ref{der-res}) review the integrability properties in the framework of the derivative  expansion, which is an alternative to the Fefferman--Graham expansion, inspired by the  black-brane paradigm and proposed in \cite{Haack:2008cp, Bhattacharyya:2008jc, Bhattacharyya:2008ji}. Our review is based on a series of papers \cite{Caldarelli:2012cm, Mukhopadhyay:2013gja}, culminating in
\cite{Gath:2015nxa}, where the complete description of integrable boundary data was developed. Here, we will particularly insist on the r\^ole played by the absence of shear in the hydrodynamic congruence, both in the integrability itself, and for designing algebraically special bulk geometries (the App. \ref{cong} is fully devoted to this important result).
We will then move (Sec. \ref{sec:PD})  to a new example, where the reconstruction of the entire Pleba\'nski--Demia\'nski  family \cite{Plebanski:1976gy} is performed and exhibits the seed for the ``$U$-duality'' transformation, mapping $(m,n)$ to $(m',n')$. Two more appendices complete the main presentation.

\section[Bulk reconstruction from boundary data]{Bulk reconstruction from boundary data\footnote{
For the benefit of the reader, we reproduce in this section the results of \cite{Gath:2015nxa} on how exact four-dimensional Einstein spacetimes can be reconstructed from boundary data. In addition, we provide the explicit proof that these geometries, with appropriate choice of boundary data, are Petrov algebraically special.}}
 \label{der-res}

\subsection{The holographic fluid in the derivative expansion}

The reconstruction of the geometry from the boundary towards the bulk can be formulated as an ADM-type Hamiltonian evolution which, as usual, requires two pieces of fundamental holographic data.  For pure gravity dynamics, one piece is the boundary metric and the other one is the energy--momentum tensor. If the boundary system is in the hydrodynamic regime, the energy--momentum tensor describes a conformal, non-perfect fluid, but this needs not be true in general for the Hamiltonian evolution scheme to hold. Irrespective of its physical interpretation, the boundary metric together with the energy--momentum tensor  allows us to reconstruct the Einstein bulk spacetime.   

The derivative expansion assumes the existence of a null   
geodesic congruence in the bulk, defining tubes that extend from the boundary inwards. 
On the boundary, this congruence translates into a timelike congruence, and the aforementioned derivative series expansion is built on increasing derivative order of this field. At the perturbative level, the fluid interpretation is applicable and the boundary timelike congruence is always identified with the boundary fluid velocity field. Beyond the perturbative framework, however, this interpretation is not faithful due to the presence of non-hydrodynamic modes in the boundary energy--momentum tensor.

As already mentioned in the introduction,  from a boundary-to-bulk perspective,  
it is unlikely that one could explicitly resum the derivative expansion -- or the Fefferman--Graham expansion alternatively; generically the bulk solution can be achieved only in a perturbative manner. 
It is remarkable, however, that given an arbitrary class of boundary metrics it is possible to set up the 
conditions it should satisfy and the energy--momentum tensor it should be accompanied with in order for an \emph{exact} dual bulk Einstein space to exist. We refer to these as integrability properties.

In the remaining of this section, we will review these integrability properties and present 
a general boundary ansatz, which allows to reconstruct almost all known Einstein spaces. The common property of all these solutions is that they are algebraically special with respect to Petrov's classification: within the proposed method, the Weyl tensor of the four-dimensional bulk is controlled from the boundary data, and turns out to be always at least of type II. Another interesting feature is that resummation generates  non-perturbative effects \emph{i.e.} non-hydrodynamic modes. We find  e.g. Robinson--Trautman solutions  \cite{Gath:2015nxa}, whose holographic dual fluid is highly far from equilibrium or the Pleba\'nski--Demia\'nski family \cite{Plebanski:1976gy}, which we will study in Sec. \ref{sec:PD}.

\subsection{The boundary data}

Consider a three-dimensional spacetime playing the r\^ole of the boundary, equipped with a metric $\text{d}s^2=g_{\mu\nu}\text{d}x^\mu \text{d}x^\nu$ ($\mu, \nu, \ldots = 0,1,2$) and with a symmetric, traceless and covariantly conserved tensor $\text{T}=T_{\mu\nu}\text{d}x^\mu \text{d}x^\nu$. We assume for this tensor the least requirements for being a conformal energy--momentum  tensor \cite{hawking1975large}, and consider systems for which it can be put in the form
\begin{equation}
\label{Tdec}
\text{T}=\text{T}_{(0)}+\Pi  
\end{equation}
with the a perfect-fluid part
\begin{equation}
\label{Tperf}
\text{T}_{(0)}=\frac{\varepsilon}{2}\left(3\text{u}^2+\text{d}s^2\right) \ .
\end{equation}
The timelike congruence $\text{u}=u_\mu(x) \text{d}x^\mu$ is normalized ($u_\mu u^\mu = -1$) and defines the fluid lines. The tensor $\Pi$ captures \emph{all} corrections to the perfect-fluid component, \emph{i.e.} hydrodynamic and non-hydrodynamic modes. The hydrodynamic part is the viscous fluid contribution, which can be expressed as a series expansion with respect to derivatives of $\text{u}$. 
The first derivatives of the velocity field are canonically decomposed in terms of the acceleration $\text{a}$, the expansion $\Theta$, the shear $\sigma$ and the vorticity (a reminder is provided in  App. \ref{rem})
\begin{equation}\label{vortgen}
\omega=\frac{1}{2}\omega_{\mu\nu }\, \mathrm{d}\mathrm{x}^\mu\wedge\mathrm{d}\mathrm{x}^\nu  =\frac{1}{2}\left(\mathrm{d}\mathrm{u} +
\mathrm{u} \wedge\mathrm{a} \right).
\end{equation}
In the Landau frame, the hydrodynamic component of $\Pi$ is transverse to $\text{u}$. The full $\Pi$ \emph{is not} transverse but 
\begin{equation}
\label{Piprop}
\Pi_{\mu\nu}u^\mu u^\nu=0 \quad \Rightarrow \quad T_{\mu\nu}u^\mu u^\nu=\varepsilon(x).
\end{equation}
The latter is the local energy density, related to the pressure via the conformal equation of state $\varepsilon=2p$. However, it should be stressed that the presence of a non-hydrodynamic component tempers the fluid interpretation. In particular, it is not an easy  task to extract the congruence $\text{u}$ from $\text{T}$, because its 
meaning as a vector tangent to fluid lines becomes questionable. 
Another important structure in three spacetime dimensions, where the Weyl tensor vanishes, is the Cotton tensor\footnote{The Cotton and Levi--Civita are pseudo--tensors, \emph{i.e.} they change sign under a parity transformation. It is therefore important  to state the convention in use.}
\begin{equation}
\label{cotdef}
C^{\mu\nu}=\eta^{\mu\rho\sigma}
\nabla_\rho \left(R^{\nu}_{\hphantom{\nu}\sigma}-\frac{R}{4}\delta^{\nu}_{\hphantom{\nu}\sigma} \right),
\end{equation}
with $\eta^{\mu\nu\sigma}=\nicefrac{\epsilon^{\mu\nu\sigma}}{\sqrt{-g}}$. This tensor vanishes if and only if the spacetime is conformally flat. It shares the key properties of the energy--momentum tensor, \emph{i.e.} it is symmetric, traceless and covariantly conserved. For later reference we introduce a contraction analogous to the energy density \eqref{Piprop}
\begin{equation}
\label{cofx}
C_{\mu \nu}u^\mu u^\nu = c(x)\ .
\end{equation}

\subsection{Bulk Petrov classification and the resummability conditions} \label{Pet}

The Weyl tensor in four dimensions can be classified according to the Petrov types. For an Einstein space this provides a complete classification of the curvature tensor.

The Petrov classification is obtained from the eigenvalue equation for the Weyl tensor. 
In particular, the Weyl tensor and its dual can be used to form a pair of complex-conjugate tensors.
Each of these tensors has two pairs of bivector indices, which can be used to form a complex two-index tensor. Its components are naturally packaged inside a complex symmetric $3 \times 3$ matrix $\text{Q}$ with zero trace (see e.g. \cite{Stephani:624239} for this construction).
This matrix encompasses the ten independent real components of the Weyl tensor and the associated eigenvalue equation determines the Petrov type.

We can now establish a connection of the bulk type with the three-dimensional boundary data. 
Performing the Fefferman--Graham expansion of the complex Weyl tensor $\text{Q}^{\pm}$ for a general Einstein space, one can show that the leading-order ($\nicefrac{1}{r^3}$) coefficient, say $\text{S}^{\pm}$, exhibits a specific combination of the components of the boundary Cotton and energy--momentum tensors.\footnote{We will provide the details in the already announced upcoming publication (see also e.g. \cite{Mansi:2008br, Mansi:2008bs}).} 
The Segre type of this combination determines precisely the Petrov type of the four-dimensional bulk metric and establishes a one-to-one map between the bulk Petrov type and the boundary data.

Assume now that we wish to reconstruct the Einstein bulk spacetime from a set of boundary data.
Given a three-dimensional boundary metric, one can \emph{impose} a desired \emph{canonical form} for the asymptotic Weyl tensor $\text{S}^{\pm}$, as e.g. a perfect-fluid form (type D) or matter--radiation form (type III or N) or a combination of both (type II) (see e.g. \cite{Chow:2009km} for these structures).
Doing so, we design from the boundary the Petrov structure of the bulk spacetime, and furthermore provide a set of conditions that turn out to guarantee the resummability of the perturbative expansion into an exact Einstein space. This procedure refers to the step one mentioned in the introduction.

It turns out that it is somehow easier to work with a different pair of complex-conjugate tensors
\begin{equation}
	\label{eqn:Tref}
	T_{\mu\nu}^\pm = T_{\mu\nu} \pm \frac{i}{8\pi G k^2 }C_{\mu\nu} \ ,
\end{equation}
{where $k$ is a constant
and $\text{T}^\pm$ is related to  $\text{S}^{\pm}$ by a similarity transformation: 
$\text{T}^\pm = \text{P} \, \text{S}^{\pm}\text{P}^{-1}$ with $\text{P}={\rm diag}(\mp i,-1,1)$.} 
Choosing a specific form for these tensors, and assuming a boundary metric $\text{d}s^2$, we are led to two conditions. The first, provides a set of equations that the boundary metric must satisfy:
\begin{equation}
\label{C-con}
\text{C}=8\pi G k^2\,  \text{Im} \text{T}^+.
    \end{equation}
The second delivers the boundary energy--momentum tensor it should be accompanied with for an exact bulk ascendent spacetime to exist:
\begin{equation}
\label{T-con}
\text{T}= \text{Re} \text{T}^+.
    \end{equation}
The tensors given in Eq.~\eqref{eqn:Tref} are by construction symmetric, traceless and conserved:
\begin{equation}
\label{Tref-cons}
\nabla\cdot \text{T}^\pm=0.
    \end{equation}
 We will refer to them as the \emph{reference energy--momentum tensors} as they play the r\^ole of a pair of fictitious conserved boundary sources, always accompanying the boundary geometry. 
It turns out that the particular combination \eqref{eqn:Tref} of the energy--momentum and Cotton tensors is exactly the combination one finds if the Weyl tensor is decomposed into self-dual and anti-self-dual components, complex-conjugate in Lorentzian signature.

Finally, we note that some care must be taken when working with $\text{T}^\pm$ instead of $\text{S}^\pm$. Indeed, the eigenvalues are equal, but not necessarily their eigenvectors. In particular, this means that one cannot determine the Petrov type unambiguously if considering the eigenvalue equation for $\text{T}^{\pm}$.\footnote{In fact, the ambiguity is only between type D and type II, since these types have the same number of eigenvalues. This was noticed e.g. in the Robinson--Trautman  metric studied in \cite{Gath:2015nxa}.}

\subsection{The derivative expansion and its resummation ansatz}

We have listed in the previous section all boundary ingredients needed for reaching holographically exact bulk Einstein spacetimes. We would like here to discuss their actual reconstruction. We will use for that the derivative expansion, organized around the derivatives of the boundary fluid velocity field $\text{u}$. This expansion assumes small derivatives, small curvature, and small higher-derivative curvature tensors for the boundary metric. This limitation is irrelevant for us since we are ultimately interested in resumming the series. A related and potentially problematic issue, is the definition of $\text{u}$, which is not automatic when the boundary energy--momentum tensor $\text{T}$ is not of the fluid type. In that case $\text{u}$ should be considered as an ingredient of the ansatz rather related to the metric than to the energy--momentum tensor and \emph{a posteriori} justified by the success of the resummation.

The guideline for the reconstruction of spacetime  based on the derivative expansion is \emph{Weyl covariance} \cite{Haack:2008cp, Bhattacharyya:2008jc}: the bulk geometry should be insensitive to a conformal rescaling of the boundary metric $\text{d}s^2\to  \nicefrac{\text{d}s^2}{{\cal B}^2}$. The latter is accompanied with $\text{C}\to {\cal B}\, \text{C}$, and at the same time $\text{T}\to {\cal B}\, \text{T}$, $\text{u}\to \nicefrac{\text{u}}{{\cal B}}$ (velocity one-form) and $\omega\to \nicefrac{\omega}{{\cal B}}$ (vorticity two-form). Covariantization with respect to rescalings requires to introduce a Weyl connection one-form:
\begin{equation}
\label{Wcon}
\text{A}:=\text{a} -\frac{\Theta}{2} \text{u} \ ,
\end{equation}
which transforms as $\text{A}\to\text{A}-\text{d}\ln {\cal B}$. Ordinary covariant derivatives $\nabla$ are thus traded for Weyl-covariant ones $\mathscr{D}=\nabla+w\,\text{A}$, $w$ being the conformal weight of the tensor under consideration. In three spacetime dimensions, Weyl-covariant quantities are e.g. 
\begin{eqnarray}
\mathscr{D}_\nu\omega^{\nu}_{\hphantom{\nu}\mu}&=&\nabla_\nu\omega^{\nu}_{\hphantom{\nu}\mu},\\
\mathscr{R}&=&R +4\nabla_\mu A^\mu- 2 A_\mu A^\mu \, \label{curlR}
\end{eqnarray}
while
\begin{equation}
\label{sigma}
\Sigma=
\Sigma_{\mu\nu} 
\text{d}x^\mu\text{d}x^\nu=-2\text{u}\mathscr{D}_\nu \omega^\nu_{\hphantom{\nu}\mu}\text{d}x^\mu- \omega_\mu^{\hphantom{\mu}\lambda} \omega^{\vphantom{\lambda}}_{\lambda\nu}\text{d}x^\mu\text{d}x^\nu
-\text{u}^2\frac{\mathscr{R}}{2} \ ,
\end{equation}
is Weyl-invariant. Notice also that for any symmetric and traceless tensor $S_{\mu\nu}\text{d}x^\mu\text{d}x^\nu$ of conformal weight $1$ (like the energy--momentum tensor and the Cotton tensor), one has
\begin{equation}
\mathscr{D}_\nu S^{\nu}_{\hphantom{\nu}\mu}=\nabla_\nu S^{\nu}_{\hphantom{\nu}\mu} \ .
\end{equation}

In the present analysis, we will be interested in situations where the boundary congruence $\text{u}$ is \emph{shear-free}. 
Despite this limitation, wide classes of dual holographic bulk geometries remain accessible. 
Vanishing shear simplifies considerably the reconstruction of the asymptotically AdS bulk geometry because it reduces the available Weyl-invariant terms. As a consequence, at each order of $\mathscr{D}\text{u}$, the terms compatible with  Weyl covariance of the bulk metric $\text{d}s^2_{\text{bulk}}$ are nicely organized. Even though we cannot write them all at arbitrary order, the structure of the first orders has suggested that resummation, whenever possible, should lead to the following \cite{Haack:2008cp, Bhattacharyya:2008jc, Bhattacharyya:2008ji, Caldarelli:2012cm, Mukhopadhyay:2013gja,
Gath:2015nxa, Petropoulos:2014yaa}:
\begin{equation}
\text{d}s^2_{\text{res.}} =
-2\text{u}(\text{d}r+r \text{A})+r^2k^2\text{d}s^2+\frac{\Sigma}{k^2}
+ \frac{\text{u}^2}{\rho^2} \left(\frac{3 T_{\lambda \mu}u^\lambda u^\mu}{k \kappa }r+\frac{C_{\lambda \mu}u^\lambda \eta^{\mu\nu\sigma}\omega_{\nu\sigma}}{2k^6}\right).
\label{papaefgenres}
\end{equation}
Here $r$ the radial coordinate whose dependence is explicit, 
$x^\mu$ are the three boundary coordinates extended to the bulk, on which depend implicitly the various functions,
$\kappa=\nicefrac{3k}{8\pi G}$, $k$ a constant,
and $\Sigma$ is displayed in \eqref{sigma}. Finally,
\begin{equation}\label{rho2}
 \rho^2=r^2 +\frac{1}{2k^4} \omega_{\mu\nu} \omega^{\mu\nu} = r^2 +\frac{q^2}{4k^4}
\end{equation}
performs the resummation as the derivative expansion is manifestly organized in powers of $q^2=2 \omega_{\mu\nu} \omega^{\mu\nu}$.
Note that the three-dimensional Hodge dual of the vorticity is always aligned with the velocity field and this is how $q(x)$ is originally defined:
\begin{equation}
\label{Hodge.vorticity}
\eta^{\mu\nu\sigma}\omega_{\nu\sigma}=q u^\mu\,.
\end{equation}
In expression \eqref{papaefgenres}, we recognize the energy density $\varepsilon(x)$ introduced in Eq. \eqref{Piprop}, and 
 $c(x)$ as in \eqref{cofx}. 
The presence of the boundary Cotton tensor stresses that the bulk is generically  asymptotically \emph{locally} anti-de Sitter. It is readily checked that boundary Weyl transformations correspond to bulk diffeomorphisms, which can be reabsorbed into a redefinition of the radial coordinate: $r\to {\cal B}\, r$.
The four-dimensional metric $\text{d}s^2_{\text{res.}}$ displayed in \eqref{papaefgenres} is not expected to be Einstein for arbitrary boundary data $\text{T}$ and $\text{d}s^2$. \emph{Our claim is that when these data satisfy Eqs. \eqref{C-con} and \eqref{T-con},  $\text{d}s^2_{\text{res.}}$ is Einstein with $\Lambda=-3 k^2$.}

More can be said on the allowed reference tensors $\text{T}^\pm$. Expression  \eqref{papaefgenres} also 
contains $\text{u}$, assumed to be the timelike and shear-free boundary hydrodynamic congruence. On the bulk 
\eqref{papaefgenres}, $\text{u}$ is a manifestly a null congruence, associated with the vector $\partial_r$. 
One can show that this congruence is \emph{geodesic and shear-free} -- the proof is displayed in App. \ref{4to3}. 
According to the Goldberg--Sachs theorem and its generalizations, the anticipated Einstein bulk metric  \eqref{papaefgenres}  is therefore 
algebraically special , 
\emph{i.e.} 
of Petrov type II, III, D, N or O. Following the discussion of Sec. \ref{Pet} on the relationship between the bulk Weyl tensor 
and the boundary reference tensors $\text{T}^\pm$ regarding Petrov classification, we conclude that the boundary piece of 
data $\text{T}^\pm$ is further constrained: it can only be of a special canonical Segre type. A boundary metric accompanied 
with a  generic $\text{T}^\pm$, even satisfying  Eqs. \eqref{C-con} and \eqref{T-con}, is not expected to guarantee  
\eqref{papaefgenres} be Einstein. Only perfect-fluid, pure matter or pure radiation (or any combination) $\text{T}^\pm$'s will 
produce an Einstein space, which will furthermore be algebraically special. Scanning over canonical forms for $\text{T}^\pm$ 
amounts to exploring various Petrov classes. \emph{Hence, Eqs.  \eqref{C-con} and \eqref{Tref-cons} appear as a boundary 
translation of Einstein's equations, in the integrable sector of algebraically special geometries.}

We would like to insist again on the r\^ole played by the absence of shear for the boundary fluid congruence, intimately related with the resummability of the derivative expansion. Not only this assumption enables to discard the large number of Weyl-covariant tensors available when the shear is non-vanishing, which would have probably spoiled any resummation attempt; but it also selects the algebraically special geometries, known to be related with integrability properties. Of course this is not a theorem and we cannot exclude that some exact Einstein type I space might be successfully reconstructed or that none exact resummation involves a congruence with shear.

We come finally to the actual definition of the boundary hydrodynamic congruence $\text{u}$. As already emphasized previously, the energy--momentum tensor $\text{T}$, obtained in the procedure described in Sec. \ref{Pet}, is not necessarily of the fluid type (we shall soon meet examples in Sec. \ref{exam} and App. \ref{emcomp}). Hence, it is not straightforward to extract from this tensor the velocity congruence $\text{u}$, required in the resummed expression \eqref{papaefgenres} -- and further check or impose the absence of shear. It seems therefore that we are led to choosing $\text{u}$ rather than determining it, as part of the ansatz of this constructive approach.

At this stage, the reader might be puzzled by the freedom in making such a choice for $\text{u}$ independently of the other boundary data such as the metric and the energy--momentum tensor. In fact, this freedom is only apparent because there is basically a unique timelike normalized and
real shearless congruence on three-dimensional geometries -- another peculiar feature of four-dimensional bulks. Indeed, given a generic three-dimensional metric, there is a unique way to express it as a fibration over a conformally flat two-dimensional base:\footnote{See e.g. \cite{Coll} and our discussion in App. \ref{4to3}. This statement is not true in the presence of isometries, where more shearless congruences may exist. In these cases, the distinct congruences are equivalent.} 
\begin{equation}
\label{PDbdymet}
\text{d}s^2=-\Omega^2(\text{d}t-\text{b})^2+\frac{2}{k^2P^2}\text{d}\zeta\text{d}\bar\zeta,
\end{equation}
with $P$ and $\Omega$ arbitrary real functions of $(t,\zeta, \bar \zeta)$, 
and\footnote{We could even set $\Omega=1$, without spoiling the generality -- as we are interested in the conformal class.} 
\begin{equation}
\label{frame}
\text{b}=B(t,\zeta, \bar \zeta)\, \text{d}\zeta+\bar B(t,\zeta, \bar \zeta)\, \text{d}\bar\zeta.
\end{equation}
In this metric, 
\begin{equation}
\label{ut}
\text{u}= -\Omega(\text{d}t-\text{b})
\end{equation}
is precisely normalized and shear-free (see App. \ref{4to3} for details).   This defines our fluid congruence, and is part of our resummation ansatz. 
Making  use of \eqref{Hodge.vorticity}
and \eqref{PDbdymet} we find 
\begin{equation}
\label{papaefgentetr}
\text{d}s^2_{\text{res.}} =-2\mathbf{k}\mathbf{l}+2\mathbf{m}\bar{\mathbf{m}},
\end{equation}
where
\begin{equation}
\label{km}
\mathbf{k}=-\text{u},\quad
\mathbf{m}=\frac{\rho}{P}\text{d}\zeta
\end{equation}
and\footnote{The Hodge duality is here meant with respect to the three-dimensional boundary:
$\ast(\text{u}\wedge \text{d} q)= \eta_{\mu}^{\hphantom{\mu}\nu\sigma}u_\nu\partial_\sigma q\, \text{d}x^\mu$.}  
\begin{equation}
\label{l}
\mathbf{l}=-\text{d}r-r \text{a} -H \text{u}
+
\frac{1}{2k^2} \ast(\text{u}\wedge (\text{d} q+q\text{a}))
\end{equation}
with
\begin{equation}
\label{Hgen}
2 H= r^2k^2 -r\, \Theta
+\frac{q^2}{k^2}
+\frac{\mathscr{R}}{2k^2}
-\frac{3 }{\rho^2 k} \left(\frac{r\varepsilon}{\kappa}+\frac{qc}{ 6k^5}
\right).
\end{equation}
In the latter expression we have introduced
 $\varepsilon(x)$ and $c(x)$ defined in \eqref{Piprop} and \eqref{cofx} ($x$ refers to the coordinates $t,\zeta,\bar\zeta$ common for bulk and boundary). The congruences $\mathbf{k}$, $\mathbf{l}$, $\mathbf{m}$ and $\bar{\mathbf{m}}$ define a null tetrad, of which $\mathbf{k}$ is geodesic and shear-free, as already stated and proven in App. \ref{4to3}.
 
 \subsection{Comments}

Several comments are in order here for making the picture complete. Equation \eqref{papaefgentetr} is obtained using the derivative expansion, which is an alternative to the Fefferman--Graham expansion and better suited for our purposes. As such, it assumes that the boundary state is in the hydrodynamic regime, described by an energy--momentum tensor of the fluid type. The latter has a natural built-in velocity field, interpreted as the fluid velocity congruence. Our method, however, does not necessarily lead to a fluid-like energy--momentum tensor. This is not a principle problem, because non-perturbative contributions with respect to the derivative expansion (non-hydrodynamic modes) are indeed expected to emerge along with a resummation \cite{Heller:2013fn}. In practice, it requires information regarding the velocity field around which the hydrodynamic modes are organized. 
Thanks to the assumption of absence of shear, crucial for eliminating many terms in the derivative expansion of the bulk metric and making it resummable, this velocity field is naturally provided by the boundary metric itself, when put in the form \eqref{PDbdymet}; it is given in Eq. \eqref{ut}.

Many examples illustrate how the method works in practice. In \cite{Gath:2015nxa} we presented the reconstruction of generic boundary data with a vorticity-free congruence. These lead to the complete family of Robinson--Trautman bulk Einstein spaces. In the following section we will consider congruences with vorticity leading to the family of Pleba\'nski--Demia\'nski.  This is important for three reasons. Firstly, the Pleba\'nski--Demia\'nski captures all aspects of black-hole physics. Secondly, the presence of vorticity encoded in $\rho^2$ (see \eqref{rho2}) makes $2H$ in \eqref{Hgen} a genuinely resummed series expansion, which for vanishing $q$ (as in Robinson--Trautman) is rather a truncated expansion. This demonstrates that we are really probing a non-trivial integrability corner of Einstein's equations. Thirdly,  the non-trivial structure behind the reference tensors $\text{T}^\pm$ turns out to open a window towards the ``$U$-duality group'' quoted in the introduction. 

\section{The Pleba\'nski--Demia\'nski Einstein spaces} \label{sec:PD}

\subsection{The boundary metric and the reference energy--momentum tensors}

The resummability of the derivative expansion, irrespective of the dimension, was originally observed in \cite{Bhattacharyya:2008jc} for the Kerr black holes. This property was latter shown to hold more systematically in four dimensions, even in the presence of a nut charge, which accounts for asymptotically locally anti-de Sitter spacetimes. This was achieved in \cite{Caldarelli:2012cm, Mukhopadhyay:2013gja, Petropoulos:2014yaa} by including an infinite, though resummable series of terms built on the boundary Cotton tensor (last term of Eq. \eqref{papaefgenres}). There, the requirement was that  the Cotton tensor of the boundary metric be proportional to the energy--momentum tensor, itself being of a perfect-fluid form. In other words, the corresponding reference tensors $\text{T}^\pm$ were proportional and  both of the Segre type D.  This kind of  ansatz turns out to unify all known black-hole solutions with mass, nut charge and rotation, which are Petrov type D. All these have two commuting Killing vectors, but they do not exhaust the Petrov-D two-Killing-vector family of Einstein spaces, known as Pleba\'nski--Demia\'nski \cite{Plebanski:1976gy} (see also \cite{PMP-GP}). The latter possess an extra parameter: the acceleration parameter -- not to be confused with the boundary fluid acceleration mentioned earlier. 

The general method for reconstructing bulk Einstein spaces described above in Sec. \ref{der-res} is based on a family of boundary metrics, equipped with a shearless congruence (supporting the fluid whenever this makes sense), together with a boundary pair of complex-conjugate  reference energy--momentum tensors, the type of which controls the Petrov type of the resummed bulk algebraically special Einstein space. The Pleba\'nski--Demia\'nski  metric is the most general Petrov-D Einstein space with two commuting Killing fields, one timelike and the other spacelike. We expect therefore to reach this metric holographically with a boundary 
possessing two isometries. 

The boundary metric will eventually be of the general type \eqref{PDbdymet} accompanied with the shearless congruence $\text{u}$ as in \eqref{ut}, and with non-vanishing $\text{b}$ in order to create vorticity \eqref{vortgen}. Indeed, as observed in \cite{Leigh:2011au, Leigh:2012jv}, boundary vorticity is necessary for generating bulk rotation and nut charge, both present in Pleba\'nski--Demia\'nski. As already mentioned, the reference tensors $\text{T}^\pm$ will be chosen of the perfect-fluid form (D-type Segre), not proportional to each-other though, for if they were we would recover the general Kerr--Taub--NUT subfamily, which misses the black-hole acceleration parameter. This latter caution forbids $\partial_t = \Omega \text{u}$ be the timelike Killing vector. Indeed, if $\partial_t$ were a Killing field, the Weyl connection \eqref{Wcon} would be exact.\footnote{The congruence $\text{u}=\nicefrac{\partial_t}{\Omega}$ is in this case shearless and expansionless with
$
\text{A}=\text{d}\ln \Omega
$.}
As a consequence, using the result of App. \ref{appendix.perfect}, the boundary energy--momentum tensor would be of the perfect-fluid form. Since $\text{T}^\pm$ are also chosen of that form, we would learn from \eqref{C-con} and \eqref{T-con} that the Cotton and the energy--momentum tensors were proportional.  

Although appropriate for a compact and elegant expression of the resummed bulk metric \eqref{papaefgentetr}, the form \eqref{PDbdymet}  for the boundary metric is not convenient for implementing the set of \emph{a priori} requirements listed above, in particular the one regarding the timelike Killing field. We will therefore parameterize differently the most general two-Killing boundary metric, adapting two coordinates $\tau$ and $\varphi$ to the two Killing commuting fields, and letting $\chi$ be the third one. Up to an arbitrary conformal factor, which plays no r\^ole in holographic issues where the important piece of data is the conformal class, such a metric can be expressed in terms of two arbitrary functions $F(\chi)$ and $G(\chi)$ as follows:
\begin{equation}
\label{PDbry}
\text{d}s^2=
-\frac{F -\chi^4 G }{F+G}
\text{d}\varphi^2 +
\frac{G -\chi^4 F }{F+G}\text{d}\tau^2
+2\chi^2\, \text{d}\varphi\, \text{d}\tau+\frac{\text{d}\chi^2}{F G }.
 \end{equation}
This expression is inspired from the boundary metric as it appears in the reconstruction of the $C$-metric, a type-D Petrov member of the general Robinson--Trautman family studied in \cite{Gath:2015nxa}, common to the Pleba\'nski--Demia\'nski one. 

We now turn to the ansatz for the reference energy--momentum tensors, chosen here of the perfect-fluid\footnote{This form of $\text{T}^\pm$ is Segre type D. Alternative choices are available, such as e.g. pure-radiation reference tensors. This option was exploited in \cite{Gath:2015nxa}, with a different boundary metric, and produced Petrov-N Robinson--Trautman Einstein spaces. Similarly,  we could proceed here with a more general ansatz aiming at recovering all possible two-Killing Einstein spaces. Restricting ourselves to the perfect-fluid form, we will only reproduce the Petrov-D ones.} form $\text{T}^\pm_{\text{pf}}$. We need for this an ansatz for two complex-conjugate, normalized congruences with
exact Weyl connection (see App. \ref{appendix.perfect}). Thanks to the presence of the two Killing fields, it is easy to design such congruences by normalizing a linear combination of these fields. Indeed, the following  
\begin{equation}
\label{upmPD}
\text{u}_{\pm}
=
\frac{\partial_\tau\pm i \partial_\varphi}{\chi^2\mp i}
\leftrightarrow
\text{u}^{\pm}=\frac{\pm i}{F+G}\left(
G\left(\text{d}\tau+\chi^2 \text{d}\varphi\right)
\pm i 
F\left(\chi^2\text{d}\tau -\text{d}\varphi\right)
\right)
 \end{equation}
are non-expanding and accelerating\footnote{They are also shearless and carry vorticity, but this plays no r\^ole here.} with
\begin{equation}
\label{apm}
\text{a}^{\pm}=
\frac{2\chi\text{d}\chi}{\chi^2\mp i},
 \end{equation}
and their Weyl connections
\begin{equation}
\label{Apm}
\text{A}^{\pm}=
\text{d}
\ln(\chi^2\mp i)
 \end{equation}
 are exact. The corresponding perfect-fluid reference tensors are then conserved with a $\chi$-dependent pressure:
\begin{equation}
\label{PD-perflu}
\text{T}^\pm_{\text{pf}}= \frac{M_\pm(\chi) k^2}{8\pi G}\left(3\left(\text{u}^\pm\right)^2 +\text{d}s^2\right),
\end{equation}
where
\begin{equation}
\label{Ppm}
M_\pm(\chi) =\frac{-m\pm i n}{(\chi^2\mp i)^3}.
\end{equation}
It is worth stressing that $m$ and $n$ are arbitrary parameters, which will survive all the way up to the bulk metric, where they will play the r\^ole of the mass and nut charge (in appropriate normalizations) respectively. 
They appear here as first integrals of \eqref{Tref-cons}. We will comment again on this point later in the discussion of Sec. \ref{Ud}.

\subsection{Cotton, integrability conditions and energy--momentum}\label{exam}

The Cotton tensor can be computed for the general boundary ansatz \eqref{PDbry}. The expressions are quite lengthy, and will not be reproduced here. The resummability condition \eqref{C-con} can be imposed using the reference tensors \eqref{PD-perflu}. They result in third-order differential equations for the functions $F(\chi)$ and $G(\chi)$. 

The equations at hand turn out to be tractable.  At the first place, their compatibility implies that 
\begin{equation}
\label{FG}
F(\chi)+G(\chi)
= k^2,
 \end{equation}
showing that the freedom on the boundary metric is severely reduced by the integrability condition \eqref{C-con}. Once this condition is imposed, some equations become linear with $\chi$-dependent coefficients and their resolution is immediate: 
\begin{equation}
\label{FGRQ}
F(\chi)=\frac{\hat R(\chi)}{\chi^4+1}, \quad G(\chi)= \frac{\hat Q(\chi)}{\chi^4+1},
\end{equation}
where $\hat Q(\chi)$ and $\hat R(\chi)$ are fourth-degree polynomials:
\begin{eqnarray}
\label{F}
\hat R(\chi)&=&(k^2-\ell)\chi^4-2n\chi^3+\epsilon \chi^2-2m\chi +\ell,\\
\label{G}
\hat Q(\chi)&=&\ell\chi^4+2n\chi^3-\epsilon \chi^2+2m\chi +k^2-\ell.
\end{eqnarray}
Two further integration constants appear: $\ell$ and $\epsilon$. Expressions \eqref{FG}, 
 \eqref{FGRQ},
 \eqref{F} and  \eqref{G}  should be thought of as integrability conditions, since they stem out of \eqref{C-con}.
 
The next step in our procedure is to determine the genuine boundary energy--momentum tensor, using \eqref{T-con}, which together with the above integrability relations is expected to guarantee that the resummed bulk metric \eqref{papaefgenres} is exact Einstein. The expression is provided in App. \ref{emcomp}, together with that of the Cotton tensor. The energy--momentum tensor  is the starting point for the physical analysis of the holographic state: it contains information on both hydrodynamic and non-hydrodynamic modes that can be extracted using, among others, the fluid congruence $\text{u}$ that will be studied in the next section. We will not perform this analysis, which stands beyond the scope of the present paper.

\subsection{The canonical frame and the hydrodynamic congruence}\label{cancon}

As stressed in the general presentation of Sec. \ref{der-res}, the hydrodynamic congruence $\text{u}$ is part of the resummation ansatz. Once the boundary metric is set in the form \eqref{PDbdymet}, the velocity field used in the resummation formula \eqref{papaefgenres} should be taken to be \eqref{ut}, which is shear-free. Our task is thus to turn \eqref{PD} onto \eqref{PDbdymet}. This is performed assuming \eqref{FG}\footnote{For the sake of clarity many expressions will still contain both $F$ and $G$, even though these are no longer independent: $F+G = k^2$. The boundary metric in particular reads:
$$
\text{d}s^2=
-\frac{F -\chi^4 G }{k^2}
\text{d}\varphi^2 +
\frac{G -\chi^4 F }{k^2}\text{d}\tau^2
+2\chi^2\, \text{d}\varphi\, \text{d}\tau+\frac{\text{d}\chi^2}{F G }.$$} (but not \eqref{FGRQ}, \eqref{F} and  \eqref{G}, even though ultimately these are necessary for the resummation to be successful).  

The coordinate change from $(\tau,\chi,\varphi)$ to $(t, \zeta, \bar\zeta)$ is easily found to be the following:
\begin{eqnarray}
\label{chi}
t+\frac{\zeta+\bar\zeta}{k\sqrt{2}}&=&\int\text{d}\chi\left(
\frac{k^2\chi^2}{FG(\chi^4+1)}-\frac{1}{F\chi^2}
\right),
\\
\label{zeta}
\zeta&=&-\frac{k}{\sqrt{2}}
\left(\tau+i\varphi-k^2\int\text{d}\chi\frac{\chi^2-i}{FG\left(\chi^4+1\right)}\right).
\end{eqnarray}
In the new frame, the two Killing vector fields read:
 \begin{equation}
\partial_{\tau}=\partial_t-\frac{k}{\sqrt{2}}
\left(\partial_\zeta+
\partial_{\bar \zeta}
\right),
\quad
\partial_{\varphi}
=-
i\frac{k}{\sqrt{2}}
\left(\partial_\zeta-
\partial_{\bar \zeta}
\right).
\label{Ktauphi}
\end{equation}
As anticipated, $\partial_t$ \emph{is not }a Killing. The boundary metric has now the form \eqref{PDbdymet}, 
\begin{equation}
\nonumber
\text{d}s^2=-\Omega^2(\text{d}t-\text{b})^2+\frac{2}{k^2P^2}\text{d}\zeta\text{d}\bar\zeta,
\end{equation}
with $\text{b}$ as in \eqref{frame},
\begin{equation}
\nonumber
\text{b}=B\, \text{d}\zeta+\bar B\, \text{d}\bar\zeta.
\end{equation}
All functions $\Omega$, $P$ and $B$ depend only on $\chi$ \emph{i.e.} on the specific combination dictated by the isometries, $t+\frac{\zeta+\bar\zeta}{k\sqrt{2}}$, via \eqref{chi}:
 \begin{eqnarray}
\label{Q}
\Omega&=&\frac{F\chi^2(\chi^4+1)}{F\chi^4-G},
\\
\label{B}
B&=&-\frac{k}{\sqrt{2}F(1+\chi^4)}\left(1-i\frac{F\chi^4-G}{k^2\chi^2}\right),
\\
\label{P}
P^2&=&\frac{k^2}{G(1+\chi^4)}.
\end{eqnarray}

We are now in position to read off the normalized hydrodynamic congruence $\text{u}$ that accompanies our ansatz of boundary metric and boundary energy--momentum tensor, as given in \eqref{ut},
\begin{equation}
\nonumber
\text{u}= -\Omega(\text{d}t-\text{b}).
\end{equation}
Expressed in the frame $(\text{d}\tau, \text{d}\chi, \text{d}\varphi)$, the latter reads:
\begin{equation}
\text{u}=\text{d}\varphi-
\chi^2\text{d}\tau +\frac{\text{d}\chi}{F}\leftrightarrow
\text{u}=\frac{F+G}{F(1+\chi^4)}\left(\chi^2\partial_\tau-\partial_\varphi\right)+G\,\partial_\chi.\label{uuPD}
 \end{equation}
Given the hydrodynamic congruence, it is possible to determine the longitudinal projections of the  energy--momentum and Cotton tensors, as displayed in Eqs. \eqref{Piprop} and \eqref{cofx}. We find:
 \begin{eqnarray}
 \label{epsPD}
\varepsilon 
(\chi)&=&-\frac{2k\kappa}{3}\frac{m\chi^2(\chi^4-3)+n(3\chi^4-1)}{(1+\chi^4)^3},\\
 \label{cPD}
c(\chi)&=&2k^4\frac{n\chi^2(\chi^4-3)-m(3\chi^4-1)}{(1+\chi^4)^3}.
 \end{eqnarray}
With  these quantities, we can rewrite $M_{\pm}(\chi)$ in \eqref{Ppm} as  
 \begin{equation}
  \label{epsMPD}
M_{\pm}(\chi)=\frac{3}{2 k \kappa}\varepsilon(\chi) \pm  \frac{i}{2 k^4}c(\chi). 
 \end{equation}
 
Besides their importance as building blocks in the resumed null tetrad \eqref{km}, \eqref{l} via \eqref{Hgen}, these expressions are interesting because they correspond to each-other (when appropriately normalized) by the following map:  $(m,n )\to (-n,m)$. This is a \emph{gravity duality map}. Gravitational duality is known to exchange the mass and the nut charge. This is settled more generally in Ricci-flat spaces, when the duality maps the Riemann tensor to its dual. As for the full continuous $U$-duality group, this $\mathbb{Z}_2$ subgroup acts non-locally on the four-dimensional metric (as opposed to the action in the three-dimensional reductions \emph{\`a la 
}); it is actually understood only in a linearized version of gravity \cite{Argurio:2009xr, Bunster:2012km}. Here  the bulk spaces are Einstein and this  $\mathbb{Z}_2$ mapping, non-local in four dimensions relates the Weyl to its dual, which on the boundary is known to exchange the Cotton and the energy--momentum tensors \cite{deHaro:2008gp, Mansi:2008br, Mansi:2008bs}. This is what we observe and  the holographic language seem to be the appropriate one for handling these duality issues in the presence of a cosmological constant. We will soon reach a similar conclusion for the generalization of the Geroch solution-generating technique to Einstein spaces.

Before moving to the resummation, let us mention here that, in the new frame,  the complex-conjugate congruences \eqref{upmPD} used in \eqref{PD-perflu} to define the perfect-fluid reference tensors  $\text{T}^\pm_{\text{pf}}$ are
 \begin{equation}
 \label{upmPDtzzb}
 \text{u}^+= \text{u}+ \frac{\alpha^+}{P^2}\text{d}\zeta, \quad
 \text{u}^-= \text{u}+ \frac{\alpha^-}{P^2}\text{d}\bar\zeta,
  \end{equation}
where 
\begin{equation}
\alpha^{\pm}
=-\frac{\nicefrac{\sqrt{2}}{k}}{\chi^2\mp i}.
\end{equation}
It is worth noting that these congruences are actually the most general ones: adding an extra leg along the missing direction ($\text{d}\bar\zeta$ or $\text{d}\zeta$ respectively), and adjusting the overall scale for keeping the normalization amounts to performing a combination of a diffeomorphism and a Weyl transformation on the metric \eqref{PDbdymet}. This shows that we are actually performing the general analysis of two-Killing boundary metrics accompanied with the most general perfect-fluid reference tensors.  Such a conclusion might not have been obvious to draw from the expressions   \eqref{upmPD}. The immediate and important consequence is that the bulk metrics we will reach by resumming the boundary data at hand, will exhaust the entire family of type-D two-Killing Einstein spaces. Other Petrov type two-Killing Einstein spaces can possibly be reconstructed, using different Segre types of reference tensors. We will not investigate this issue here. 

\subsection{From the boundary to the bulk}

At this stage we are ready to pursue with the resummation of the derivative expansion associated to the above boundary data, using Eq. \eqref{papaefgentetr}. This requires to determine the null tetrad $\mathbf{k}, \mathbf{l}, \mathbf{m}$ and $\bar{\mathbf{m}}$, using \eqref{km} and \eqref{l}.  Further intermediate ingredients are necessary for that, which we will compute and present in the frame  $(\text{d}\tau, \text{d}\chi, \text{d}\varphi)$. The bulk frame is thus $(\text{d}r,\text{d}\tau, \text{d}\chi, \text{d}\varphi)$.

The hydrodynamic boundary congruence $\text{u}$ given in \eqref{uuPD} is shearless but has acceleration, expansion and vorticity:
 \begin{eqnarray}
\label{aPD}
\text{a}&=& - 2\chi\left(G \text{d}\tau 
-
\frac{k^2\chi^2}{F(1+\chi^4)}
\text{d}\chi
\right),
 \\
 \label{thPD}
\Theta&=&\frac{4\chi^3G}{1+\chi^4}+G',
\\
\omega&=& \chi\left(
\frac{k^2}{F(1+\chi^4)}
\left(\text{d}\tau+\chi^2\text{d}\varphi\right)\wedge \text{d}\chi
-G\text{d}\varphi\wedge\text{d}\tau\right).
 \end{eqnarray}
 The dual vorticity, computed using \eqref{Hodge.vorticity} reads:
 \begin{equation}
 \label{qPD}
q=
\frac{2k^2 \chi}{1+\chi^4},
 \end{equation}
and this allows to express the resummation variable $\rho(r,\chi)$ defined in \eqref{rho2} as:
 \begin{equation}
 \label{rhoPD}
 \rho^2=r^2 + \frac{\chi^2}{\left(1+\chi^4\right)^2}.
 \end{equation}
\vskip0.2cm
Using the above, we can compute explicitly the null tetrad (Eqs. \eqref{km} and\eqref{l}), and reconstruct the bulk metric as in Eq. \eqref{papaefgentetr}. In their original form, all these expressions are not very illuminating and we recollect them in App. \ref{null}.
It is possible, however, to abandon the analogue of Eddington--Finkelstein coordinates $(\tau, \chi, \varphi, r)$ in use, and bring the metric at hand in a more familiar form. This is achieved with the following coordinate transformation:
\begin{equation}
\label{trans}
\begin{cases}
\tau&=\hat \tau+\int \frac{\hat{q}^2\, \text{d}\hat q}{\hat R(\hat q)}-\int 
\frac{\hat{r}^2\, \text{d}\hat r}{\hat R(\hat r)}
\\
\chi&=\hat q
\\
\varphi&=\hat \varphi-\int \frac{\text{d}\hat q}{\hat R(\hat q)}+\int 
\frac{\text{d}\hat r}{\hat R(\hat r)}
\\
r&=\frac{1}{\hat r-\hat q}+\frac{\hat{q}^3}{\hat{q}^4+1}
,
\end{cases}
\end{equation}
where the function $\hat R$ is displayed in \eqref{F}. Despite the complicated expression of the original metric, 
thanks to the polynomial structure of $\hat R$ and to Eqs. \eqref{FGRQ}, \eqref{F} and \eqref{G}, the metric  \eqref{papaefgentetr} becomes unexpectedly simple  in
the new coordinates $(\hat\tau, \hat q, \hat\varphi, \hat r)$:
\begin{eqnarray}
\text{d}s^2&=&\frac{1}{(\hat{q}-\hat{r})^2}\bigg(
-\frac{\hat R(\hat{r}) \left(\text{d}\hat{\varphi}-\hat{q}^2\text{d}\hat{\tau}\right)^2}{1+(\hat{r}\hat{q})^2}
+\frac{\hat Q(\hat{q}) \left(\text{d}\hat{\tau}+\hat{r}^2\text{d}\hat{\varphi}\right)^2}{1+(\hat{r}\hat{q})^2}
\nonumber
\\
\label{PD}
&&+
\frac{1+(\hat{r}\hat{q})^2}{\hat Q(\hat{q})} \text{d}\hat{q}^2
+
\frac{1+(\hat{r}\hat{q})^2}{\hat R(\hat{r})} \text{d}\hat{r}^2\bigg).
 \end{eqnarray}
This metric is known as the Pleba\'nski--Demia\'nski  geometry  \cite{Plebanski:1976gy} (see also \cite{PMP-GP}). 

The Pleba\'nski--Demia\'nski solution is the most general Einstein space with two commuting Killing vector fields.
 It is Petrov D because it possesses two multiplicity-two principal null directions.\footnote{A principal null direction $\mathbf{n}$ obeys 
 $n_{[\rho}W_{\kappa]\lambda\mu[\nu}n_{\sigma]}n^\lambda n^\mu=0$, 
 where $W_{\kappa\lambda\mu\nu}$ are the components of the Weyl tensor.} The congruence  tangent to $\mathbf{k}$ is a principal null direction, but this property does not hold for $\mathbf{l}$.
%
%
An alternative null tetrad\footnote{The null tetrad is not uniquely defined: one can perform boosts in the $\mathbf{k}$-$\mathbf{l}$ plane, null rotations (with $\mathbf{k}$ fixed) and spatial rotations in the $\mathbf{m}$-$\bar{\mathbf{m}}$ plane \cite{Stephani:624239, PMP-GP}.} is read off directly from the expression \eqref{PD}:
\begin{equation}
\label{PDprime}
 \begin{cases}
\mathbf{k}'&=\frac{1}{\sqrt{2}(\hat{q}-\hat{r})}
\left(
\sqrt{\frac{\hat R(\hat{r}) }{1+(\hat{r}\hat{q})^2}}\left(\text{d}\hat{\varphi}-\hat{q}^2\text{d}\hat{\tau}\right)
+ \sqrt{\frac{1+(\hat{r}\hat{q})^2}{\hat R(\hat{r}) }}\text{d}\hat{r}
\right)
\\
\mathbf{l}'&=\frac{1}{\sqrt{2}(\hat{q}-\hat{r})}
\left(
\sqrt{\frac{\hat R(\hat{r}) }{1+(\hat{r}\hat{q})^2}}\left(\text{d}\hat{\varphi}-\hat{q}^2\text{d}\hat{\tau}\right)
- \sqrt{\frac{1+(\hat{r}\hat{q})^2}{\hat R(\hat{r}) }}\text{d}\hat{r}
\right)
\\
\mathbf{m}'&=\frac{1}{\sqrt{2}(\hat{q}-\hat{r})}
\left(
\sqrt{\frac{\hat Q(\hat{r}) }{1+(\hat{r}\hat{q})^2}}\left(\text{d}\hat{\tau}+\hat{r}^2\text{d}\hat{\varphi}\right)
+i \sqrt{\frac{1+(\hat{r}\hat{q})^2}{\hat Q(\hat{r}) }}\text{d}\hat{q}
\right).
\end{cases}
 \end{equation}
Two of its elements, $\mathbf{k}'$ (which is actually proportional to $\text{u}=-\mathbf{k}$) and $\mathbf{l}'$ turn out to 
provide the two principal null directions, and are furthermore geodesic and shear-free (details can be found e.g. in Ref. \cite{PMP-GP, Podolsky:2009ag}). In the latter tetrad the only non-vanishing Weyl invariant is    
\begin{equation}
\Psi_2=W^{\kappa\lambda\mu\nu}\,k'_\kappa m'_\lambda \bar m'_\mu l'_\nu =(m-in)\left(
\frac{\hat{q}-\hat{r}}{i-\hat{r}\hat{q}}
\right)^3.
 \end{equation}
 Using the transformation \eqref{trans} (and \eqref{Ppm}, \eqref{epsMPD}) we find for the leading large-$r$ behaviour:
 \begin{equation}
\Psi_2\approx
-\frac{M_+(\chi)}{r^3}=-\frac{1}{r^3}\left(
\frac{3}{2 k \kappa}\varepsilon(\chi) +  \frac{i}{2 k^4}c(\chi)\right).
 \end{equation}
This last expression illustrates what has been explained in general terms in Sec. \ref{Pet}, regarding the boundary information carried by the bulk Weyl tensor. Here, for Petrov-D spaces, $\Psi_2$ (\emph{i.e.} the projection of the Weyl onto $\mathbf{k}'\propto \text{u}$) contains the information on $\varepsilon$ and $c$, which are the projections of the boundary energy--momentum and Cotton tensors onto the hydrodynamic congruence $\text{u}$.

\boldmath
\subsection{Towards a $U$-duality group}\label{Ud}
\unboldmath

We would like to stress that the Einstein space \eqref{PD} has been found here from \emph{purely boundary considerations}: (\romannumeral1) a general two-Killing boundary metric \eqref{PDbry}
with its associated normalized and shear-free congruence \eqref{uuPD}, and
(\romannumeral2) a general perfect-fluid reference tensor \eqref{PD-perflu}.  The integrability conditions \eqref{C-con} involving the boundary Cotton tensor completely determine the \emph{a priori} arbitrary functions $F$ and $G$ appearing in the metric, whereas the boundary energy--momentum tensor is determined via the other integrability condition \eqref{T-con}.
The resulting data allow to reconstruct an exact Einstein space thanks to the resummation formula \eqref{papaefgentetr}.
 
The Pleba\'nski--Demia\'nski family obtained in this way is described in terms of the four parameters entering the functions $F$ and $G$, Eqs. \eqref{FGRQ}, \eqref{F} and \eqref{G}: $\epsilon, \ell, m$ and $n$. Of these, $m$ and $n$ appear in the expression for the reference tensors  $\text{T}^\pm$, chosen here to be of the perfect-fluid type (type D in Segre classification), whereas the others, $\epsilon$ and $\ell$ emerge after solving the integrability conditions \eqref{C-con}. The two parameters $m$ and $n$ can be thought of as constants of motion of the equation \eqref{Tref-cons}, 
\begin{equation}
\nonumber
\nabla\cdot \text{T}^\pm=0,
\end{equation} 
reflecting its invariance under transformations 
\begin{equation}
\label{U-dual}
\text{T}^{\pm}
\to \text{e}^{\pm i\lambda}\, 
\text{T}^{\pm}, \ \lambda\in \mathbb{R}.
\end{equation}
Indeed, with a perfect-fluid reference tensor $\text{T}^{\pm}_{\text{pf}}$ as given in \eqref{PD-perflu} and \eqref{Ppm}, the transformation \eqref{U-dual} produces the continuous $U(1)$ mapping 
\begin{equation}
\label{U-mn}
m-in\to m'-in'=
 \text{e}^{i\lambda}\, (m-in). 
\end{equation}
For $\lambda=\nicefrac{-\pi}{2}$ in particular, this amounts to the already quoted (see Sec. \ref{cancon}) discrete $\mathbb{Z}_2$ transformation $(m,n) \to (-n,m)$, exchanging the r\^oles of $\varepsilon(\chi)$ and $c(\chi)$. Given a solution with parameters $m$ and $n$, another solution is made available with 
$(m',n')$, and this is reminiscent of the $U(1)$ subgroup of Geroch's $SL(2,\mathbb{R})$ action in Ricci-flat spaces \cite{Geroch, Geroch:1972yt}. The $\mathbb{Z}_2$ subgroup is itself the gravitational Weyl duality action, reminiscent of Geroch's Riemann duality discrete subgroup for the Ricci-flat case.

The Geroch group emerges as a symmetry of the three-dimensional sigma model obtained when reducing four-dimensional vacuum Einstein's equations along a Killing congruence. Here, a similar symmetry appears in the reference tensor equations of motion on the conformal boundary of an Einstein space. In both cases, this symmetry acts locally on the three-dimensional data, whereas the action is non-local on the four-dimensional solution itself. In practice, within the class at hand, this
this action (at least for a $U(1)$ subgroup of it) induces a transformation mixing the mass and nut charge.\footnote{Here, the parameter $n$ is not exactly the nut charge, but is closely related to it  (and dressed with the cosmological constant \emph{i.e.} $k$). The physical parameters such as the nut charge, the angular velocity or the black-hole acceleration are obtained via a rescaling accompanied with a coordinate transformation, leading to more involved expressions, which we have avoided here. These expressions can be found in \cite{PMP-GP}; they are useful because they allow to consider the limit of vanishing acceleration parameter, where one recovers the results of Ref. \cite{Mukhopadhyay:2013gja}:
$$
\text{u}^+=\text{u}^-=\text{u} \quad \text{and} \quad
\text{T}^{+}\propto \text{T}^{-}\propto \text{T}\propto \text{C}.
$$} Not only this shows that we are on the right tracks for understanding the generalization of Geroch's method to Einstein spaces, but it also gives us confidence when interpreting Eqs. \eqref{C-con},  \eqref{T-con} and \eqref{Tref-cons} as Einstein's equations with cosmological constant in some integrable sector -- much like Geroch's sigma-model equations are vacuum Einstein's equations for reductions along Killing congruences.

\section*{Conclusions}
\addcontentsline{toc}{section}{Conclusions}

The developments we have presented here are twofold. 

They concern at the first place the integrability of Einstein's equations from a holographic perspective, setting the conditions that a given class of boundary metrics should satisfy, and determining the energy--momentum tensor it should be accompanied with in order for an \emph{exact} dual bulk Einstein space to exist. The proposed procedure is three-step:
\begin{itemize}
\item The first step consists in choosing a set of two complex-conjugate reference tensors $\text{T}^\pm$, symmetric, traceless and satisfying the conservation equation \eqref{Tref-cons}.
\item Next, this tensor enables us  (\romannumeral1) to set conditions on the boundary metric by imposing its Cotton be the imaginary part of $\text{T}^\pm$ (up to constants), Eq. \eqref{C-con}; (\romannumeral2) to determine the boundary energy--momentum tensor as its real part, Eq. \eqref{T-con}.
\item Finally, using these data and Eq. \eqref{papaefgentetr}, we reconstruct the bulk Einstein space.
\end{itemize}
It is fair to say that Eqs. \eqref{C-con},  \eqref{T-con} and \eqref{Tref-cons} emerge as Einstein's equations in some integrable sector. Moreover, this sector is that of Petrov algebraically special solutions. 

Next, this procedure was applied for reconstructing the full Pleba\'nski--Demia\'nski family. The starting point is a general boundary metric with two commuting Killing vectors and perfect-fluid boundary reference tensors with complex conjugate congruences $\text{u}^{\pm}$ obeying  $\text{d}\text{A}^{\pm}=0$. This equation of motion for the reference perfect fluids is equivalent to the conservation requirement for $\text{T}^{\pm}_{\text{pf}}$, and is invariant under the transformation \eqref{U-dual}.
This transformation on the boundary data maps a bulk exact Einstein space onto another. Hence, it is solution-generating.

The above results are encouraging regarding our original motivations and expectations. They deserve further investigation. 

On the one hand, the resummability conditions for the derivative expansion presented here and borrowed from \cite{Gath:2015nxa}
are of constructive value, as we have not proven that they are necessary or sufficient: they work efficiently and systematically, 
and can generate all known algebraically special Einstein spaces. Here we discussed the Pleba\'nski--Demia\'nski familly, whereas 
the Robinson--Trautman family was explicitly built in \cite{Gath:2015nxa}. The proof that \eqref{papaefgenres} is indeed Einstein 
will appear in a future work, but some further questions remain. Our working assumption was the absence of shear for the boundary 
hydrodynamic congruence. This makes the resummation of the derivative expansion possible, and as a corollary,  implies the existence 
of a bulk null, geodesic and shear-free congruence -- Goldberg--Sachs theorem for Einstein spaces is at work. Is boundary shear a genuine obstruction to 
resummability? In our approach the absence of shear guarantees the bulk be algebraically special, and a precise relationship is set 
between the Segre type of the reference tensor and the Petrov type of the bulk Weyl tensor. This was illustrated in the case of the 
 Pleba\'nski--Demia\'nski familly, which is Petrov D and is indeed built with perfect-fluid type reference boundary tensors \emph{i.e.} 
of Segre type D. 
Can one reconstruct exact Einstein spaces which are not algebraically special, with non-zero shear on the boundary? Can one better understand the interplay between the two perturbative expansions mentioned here, namely the Fefferman--Graham and the derivative ones? 

On the other hand, the emergence of a ``holographic $U$-duality symmetry'' remains modest. Although, the observed $U(1)$ invariance is valid, of course, as is its effect on the mass and nut parameters (in agreement with our expectations inferred from the Ricci-flat paradigm),  it is at the present stage  confined to the somehow restricted boundary framework of the Pleba\'nski--Demia\'nski family. We nevertheless believe that the whole approach, new and original, starts shedding light on the integrability and solution-generating properties of Einstein's equations in the presence of cosmological constant.

\section*{Acknowledgements}

This work is based partly on a talk delivered during the workshop in honour of Prof. Ph.~Spindel \textsl{About various kinds of interactions} held in June 2015 at the University of Mons, Belgium. 
It relies on works published or to appear, performed in collaboration with M. Caldarelli, J.~Gath, R. Leigh, A. Mukhopadhyay, A. Petkou and V. Pozzoli. 

We are particularly grateful to D. Klemm for drawing in 2013 our attention to the black-hole acceleration parameter of the  
Pleba\'nski--Demia\'nski family, missing in our article \cite{Mukhopadhyay:2013gja}. This parameter turns out to be the 
cornerstone for understanding the holographic origin of integrability in Einstein spaces. We also had interesting correspondence 
with M.~Blau and J.~Podolsk\' y.

We would like to thank the organizers T. Basile, N. Boulanger, S. Detournay, M. Henneaux and I. Van Geet for this unforgettable meeting. The research of P.M. Petropoulos and K.~Siampos is supported  by the Franco--Swiss bilateral Hubert Curien program  \textsl{Germaine de Stael} 2015 (project no 32753SG). 

Finally we acknowledge each others home institutions, the National and Kapodistrian University of Athens
and the Aristotle University of Thessaloniki for hospitality and financial support, where part of this work was developed.

\appendix
\section{On congruences}\label{cong}

\subsection{General congruences}\label{rem}

Consider a $D$-dimensional hyperbolic geometry equipped with a metric 
\begin{equation}\label{Dmet}
\text{d}s^2 =g_{MN}\,\text{d}\text{x}^M \text{d}\text{x}^N
\end{equation}
with $M,N, \ldots =0,1,\ldots, D-1$ and $I,J, \ldots =1,\ldots, D-1$. We do not make any assumption regarding isometries. Consider now an arbitrary timelike vector field $\text{u}$, normalized as  $u^M u_M=-1$, and let $U$ be the longitudinal projector and $\Delta$ the projector on the locally orthogonal hyperplane:
\begin{equation}
\label{proj}
U_{MN} = - u_M u_N , \quad \Delta_{MN } =  u_M u_N  + g_{MN}.
\end{equation}
These projectors satisfy the usual identities:
\begin{equation}
\label{projprop}
U^M_{\hphantom{M}R } U^R _{\hphantom{R }N } = U^M_{\hphantom{M}N },\quad U^M_{\hphantom{M}R } \Delta^R _{\hphantom{R }N }  =   0 , \quad \Delta^M_{\hphantom{M}R } \Delta^R _{\hphantom{R }N }  =   \Delta^M_{\hphantom{M}N } ,\quad U^M_{\hphantom{M}M}=1, \quad \Delta^M_{\hphantom{M}M}=D-1.
\end{equation}

The integral lines of $\text{u}$ define a congruence characterized by its acceleration, shear, expansion and vorticity:
\begin{equation}
\label{def1}
\nabla_{M} u_N=-u_M a_N +\frac{1}{D-1}\Theta \Delta_{MN}+\sigma_{MN} +\omega_{MN}
\end{equation}
with\footnote{Our conventions are: $A_{(MN)}:=\nicefrac{1}{2}\left(A_{MN}+A_{NM}\right)$ and $A_{[MN]}:=\nicefrac{1}{2}\left(A_{MN}-A_{NM}\right)$.}
\begin{equation}
\label{def2}
\begin{cases}
a_M&=u^N\nabla_N u_M, \quad
\Theta=\nabla_M u^M \\
\sigma_{MN }&=\frac{1}{2} \Delta_M^{\hphantom{M}R } \Delta_N ^{\hphantom{N }S}\left(
\nabla_R  u_S +\nabla_S  u_R 
\right)-\frac{1}{D-1} \Delta_{MN }\Delta^{R S } \nabla_R  u_S  \\
&= \nabla_{(M} u_{N )} + a_{(M} u_{N )} -\frac{1}{D-1} \Delta_{MN } \nabla_R  u^R  
\\
\omega_{MN }&=\frac{1}{2} \Delta_M^{\hphantom{M}R } \Delta_N ^{\hphantom{N }S }\left(
\nabla_R  u_S -\nabla_S  u_R 
\right)= \nabla_{[M} u_{N ]} + u_{[M} a_{N ]}.
\end{cases}
\end{equation}
By construction, all these tensors are transverse; they satisfy the following identities:
\begin{equation}
u^M a_M=0, \quad u^M \sigma_{MN }=0,\quad u^M \omega_{MN }=0, \quad u^M \nabla_N  u_M=0, \quad \Delta^R _{\hphantom{R }M} \nabla_N  u_R  =\nabla_N  u_M.
\end{equation}

The vorticity allows to define the following form
\begin{equation}\label{def3}
2\omega=\omega_{MN}\, \mathrm{d}\mathrm{x}^M\wedge\mathrm{d}\mathrm{x}^N  =\mathrm{d}\mathrm{u} +
\mathrm{u} \wedge\mathrm{a}\, .
\end{equation}
When $\omega$ is non-closed, the field $\text{u}$ is not hypersurface-orthogonal. If $t$ is the coordinate adapted to the congruence,  
so that $\text{u}=\nicefrac{\partial_t}{\Omega}$, the corresponding form reads generally:
\begin{equation}
\label{uform}
\text{u}= -\Omega(\text{d}t-\text{b}), \quad \text{b}=b_I\,\text{d}x^I\,.
\end{equation}
With this choice of coordinates, due to \eqref{projprop}, $\Delta_{0M}=0$. Consequently only  $
\Delta_{IJ}$ are non-vanishing. Hence the metric reads:
\begin{equation}
\label{adfr}
\text{d}s^2=-\Omega^2(\text{d}t-\text{b})^2+\Delta_{IJ}\,\mathrm{d}x^I \mathrm{d}x^J.
\end{equation}

We can compute the various properties of the congruence 
 $\text{u}$ in the adapted frame at hand \eqref{adfr}. We find:
 \begin{eqnarray}
\text{a}&=&\frac1\Omega\partial_t \text{u}+\mathrm{d}\ln\Omega\,, \\
\omega&=&\frac{1}{2}\left(\Omega\, \text{db}+\frac1\Omega\text{u}\wedge\partial_t \text{u}\right),
\\
\Theta&=&\frac{1}{2\Omega}\partial_t\ln\det\Delta_{D-1}
\,,
\end{eqnarray}
where $\Delta_{D-1}$ stands for the restricted matrix of rank $D-1$.  Finally, the components of the shear tensor read:
\begin{equation}
\label{shear3d}
\sigma_{MN}=\frac\Omega2\left(\partial_t \Delta_{MN}-\frac{\Delta_{MN}}{D-1}\partial_t \ln\det\Delta_{D-1}\right)
\end{equation}
with vanishing longitudinal components $\sigma_{0M}$.

\subsection{Boundary vs. bulk shearless congruences}\label{4to3}
 
Let us now specialize to the case $D=3$ ($\mu,\nu, \ldots =0,1, 2$ and $i,j, \ldots =1,2$).
From Eq. \eqref{shear3d}, we learn that there are two obvious instances where the shear of $\text{u}=\nicefrac{\partial_t}{\Omega}$ vanishes.
\begin{enumerate}
\item The shear of $\text{u}$ vanishes when $\partial_t$ is a Killing field (cancellation of each term in \eqref{shear3d}) -- this holds actually in any dimension $D$.
\item It also vanishes when $\Delta_{ij}\,\mathrm{d}x^i \mathrm{d}x^j$ in \eqref{adfr} defines a conformally flat two-surface:
\begin{equation}
\label{adfr2-CF}
\Delta_{ij}\,\mathrm{d}x^i \mathrm{d}x^j=\frac{2}{k^2P^2}\text{d}\zeta\text{d}\bar\zeta
\end{equation}
with $P=P(t,\zeta,\bar \zeta)$ a real function (cancellation between the two terms in \eqref{shear3d}). 
\end{enumerate}
It should be clear that given an arbitrary congruence $\text{u}$, it is not always possible to bring the metric into the fibration form \eqref{adfr} with \eqref{adfr2-CF} -- if this were true, every three-dimensional congruence would be shearless. Conversely, in three dimensions, \emph{it is always possible to find a frame where the metric is a fibration over a conformally flat two-surface}, as in \eqref{adfr}, \eqref{adfr2-CF}, and this frame is generically unique.\footnote{A discussion on this issue can be found in Ref. \cite{Coll}.} The timelike congruence $\text{u}=-\Omega(\text{d}t-\text{b})$ on which this frame is adapted is thus shear-free. Therefore, irrespective of any symmetry, there is always a generically unique timelike normalized shearless congruence in $D=3$.

We now move to four dimensions ($D=4$ with indices $M = (r,\mu)$). Our scope to prove, using the above results, that the bulk velocity field $\partial_r$ of our four-dimensional resummed metric \eqref{papaefgentetr} is null, geodesic and shear-free. 
\begin{enumerate}
\item
The congruence $\partial_r$  is null as our four-dimensional metric \eqref{papaefgenres} is written in 
an analogue of Eddington--Finkelstein coordinates, \emph{i.e.} $g_{rr}=0$.
\item
The congruence $\partial_r$ is geodesic because we can easily show that its acceleration vanishes:
\begin{equation}
a^M=u^N\nabla_N u^M=\Gamma^M_{\hphantom{M}rr}=g^{M\nu}\partial_r g_{r\nu}=-g^{M\nu}\partial_r u_\nu=0\, ,
\end{equation}
since the velocity form \eqref{ut} is independent of $r$.
\item
This null and geodesic congruence $\partial_r$  turns out to be shearless. 
Consider the associated form $\mathbf{k}=-\text{u}$ together with another null field  $\mathbf{l}$, and 
define a rank-2 projector on the locally orthogonal hyperplane:
\begin{equation}
\label{null.projector}
\gamma_{MN}:=g_{MN}+k_M l_N+k_N l_M\,,\quad \mathbf{k}^2=\mathbf{l}^2=0\,,\quad \mathbf{k}\cdot\mathbf{l}=-1\,.
\end{equation}
This projector has the following properties:
\begin{equation}
\label{null.projector1}
l^M\gamma_{MN}=k^M\gamma_{MN}=0\,,\quad
 \gamma_{MP}\gamma^{PN}=\gamma_M^{\hphantom{M}N}\,,
\end{equation}
where $\gamma^{MN}:=g^{MP}\gamma_{PQ}g^{QN}$.
Using the latter we can project the covariant derivative of the congruence as follows:
\begin{eqnarray}
b_{MN}&:=&\gamma_M^{\hphantom{M}P}\gamma_N^{\hphantom{M}Q}\,\nabla_{P}\,u_Q\\
&=&
\frac12\left(b_{MN}+b_{NM}-\Theta\,\gamma_{MN}\right)+\frac12\left(b_{MN}-b_{NM}\right)+
\frac{\Theta}{2}\,\gamma_{MN}\,,
\label{null.decom}
\end{eqnarray}
where 
\begin{equation}
\Theta:=\gamma^{MN}\,b_{MN}.
\end{equation}
Expression \eqref{null.decom}
defines the shear, the vorticity and the expansion for a null geodecic congruence.
To compute the shear of the congruence $\partial_r$ we note that
\begin{equation}
\nabla_M u_N+\nabla_N u_M=\partial_r g_{MN}\,,
\end{equation}
which, thanks to \eqref{null.projector1} and \eqref{null.projector}, leads to
\begin{equation}
b_{MN}+b_{NM}=\gamma_M^{\hphantom{M}P}\gamma_N^{\hphantom{M}Q}\,\partial_r g_{PQ}=
\gamma_M^{\hphantom{M}P}\gamma_N^{\hphantom{M}Q}\,\partial_r \gamma_{PQ}\,.
\end{equation}
 In addition, the expansion can be rewritten as:
\begin{equation}
\Theta=\gamma^{MN}\gamma_M^{\hphantom{M}P}\gamma_N^{\hphantom{M}Q}\nabla_P u_Q=
\gamma^{PQ}\nabla_P u_Q=\nabla^P u_P=
\frac12\partial_r\ln g
\end{equation}
($g:=\vert\det\text{g}\vert$), 
where we used \eqref{null.projector}, \eqref{null.projector1} and the fact that $\mathbf{k}$ is null and geodesic. The components of the shear tensor are given by 
\begin{equation}
\label{shear.4d}
\sigma_{MN}=\frac12\,\gamma_M^{\hphantom{M}P}\gamma_N^{\hphantom{M}Q}
\left(\partial_r\gamma_{PQ}-\frac{\gamma_{PQ}}{2}\partial_r\ln g\right),
\end{equation}
of which the null ones are all  vanishing by construction.
Using \eqref{papaefgentetr}, \eqref{km} and \eqref{l}, we can compute the determinant of the metric 
(in the frame $(\text{d}r, \text{d}t, \text{d}\zeta, \text{d}\bar \zeta)$): 
\begin{equation}
\label{det}
g=\frac{\Omega^2\rho^4}{P^4}\,.
\end{equation}
Furthermore from \eqref{null.projector} and \eqref{papaefgentetr}  we find:
\begin{equation}
\label{projector.null}
\gamma_{PQ}\,\text{d}x^P\text{d}x^Q=2\mathbf{m}\bar{\mathbf{m}}=\frac{2\rho^2}{P^2}\text{d}\zeta\text{d}\bar\zeta\,.
\end{equation}
This is the metric on a conformally flat two-surface, and this structure is inherited from the form \eqref{PDbdymet} of the boundary metric itself, due to the shearlessness of the boundary congruence $\text{u}$, along the lines of \eqref{adfr2-CF}. With \eqref{det} and \eqref{projector.null}, and with  $\Omega$ being $r$-independent, the expression \eqref{shear.4d} for the shear of $\partial_r$ vanishes by cancellation between the two terms -- exactly as it happens for $\nicefrac{\partial_t}{\Omega}$ in three dimensions.  Had not $\nicefrac{\partial_t}{\Omega}$ been shear-free in three-dimensions, $\gamma_{PQ}\,\text{d}x^P\text{d}x^Q$ would not have been conformally flat, and $\partial_r$ would have had shear.
\end{enumerate}

Finally we recapitulate the above results, regarding the original motivations:  
{\it
An arbitrary three-dimensional Lorentzian boundary metric can be uniquely expressed as a fiber bundle
spanned by a timelike shearless vector field over a conformally flat two-dimensional base. 
The absence of shear for this boundary congruence guarantees that the corresponding null and geodesic bulk congruence is also shear-free. Thanks to the Goldberg--Sachs theorem and its generalizations, a reconstructed Einstein bulk geometry \eqref{papaefgenres}
is algebraically special.}

\subsection{Perfect-fluid congruences}
\label{appendix.perfect}

What makes a congruence $\text{u}$ be the velocity of a perfect conformal fluid? Conformal perfect fluids obey Euler's equations, which in $D$ spacetime dimensions read ($x$ stands for generic coordinates):
\begin{equation}
\label{PMP-Euler0}
 \begin{cases}
(D-1)\text{u}(\ln p)+D\, \Theta=0 \\
 \text{u}(\ln p)\, \text{u}
+\text{d}\ln p+D\, \mathrm{a}=0
\end{cases}
\end{equation}
with $p(x)$ the pressure field and
$\text{u}(f)=u^\mu \partial_\mu f$.
Combining these equations, we obtain:
\begin{equation}\label{Euler0-int}
\text{A}+\text{d}\ln p^{\nicefrac{1}{D}}=0,
\end{equation}
where 
\begin{equation}\label{WconD}
\text{A}=\text{a} -\frac{\Theta}{D-1} \text{u}.
\end{equation}
Equation \eqref{Euler0-int} is integrable if the Weyl connection $\text{A}$ is closed (hence locally exact). 
If $\text{dA}\neq 0$ the fluid flowing on $\text{u}$ is not perfect. If $\text{A}$ vanishes, the fluid is perfect
and isobar.

\section{The reconstructed boundary tensors for Pleba\'nski--Demia\'nski}\label{emcomp}

We provide in this appendix the expressions for the boundary energy--momentum tensor and for the boundary Cotton tensor, as they are obtained from Eqs. \eqref{T-con} and  \eqref{C-con}, when a perfect-fluid reference tensor \eqref{PD-perflu} is considered. We assume here that the integrability conditions \eqref{FG}, \eqref{FGRQ}, \eqref{F} and \eqref{G}, imposed on the metric by the form  \eqref{C-con} of the Cotton, are fulfilled. We find
\begin{eqnarray}
\text{T}&=&\frac{\varepsilon}{2}\text{d}s^2+\frac{3\varepsilon}{2k^4}\left(F^2\left(\text{d}\varphi-\chi^2\, \text{d}\tau\right)^2-G^2\left(\text{d}\tau+\chi^2\, \text{d}\varphi\right)^2\right)\nonumber\\
&&-\frac{\kappa c}{k^7}\,FG\left(\text{d}\varphi-\chi^2\, \text{d}\tau\right)\left(\text{d}\tau+\chi^2\, \text{d}\varphi\right)
\label{fullT}
\end{eqnarray}
and
\begin{eqnarray}
\text{C}&=&\frac{c}{2}\text{d}s^2+\frac{3c}{2k^4}\left(F^2\left(\text{d}\varphi-\chi^2\, \text{d}\tau\right)^2-G^2\left(\text{d}\tau+\chi^2\, \text{d}\varphi\right)^2\right)\nonumber\\
&&+\frac{9\varepsilon}{\kappa k}\,FG\left(\text{d}\varphi-\chi^2\, \text{d}\tau\right)\left(\text{d}\tau+\chi^2\, \text{d}\varphi\right),
\label{fullC}
\end{eqnarray}
where the functions $\varepsilon(\chi)$ and $c(\chi)$ are given in \eqref{epsPD} and \eqref{cPD}, while $F(\chi)$ and $G(\chi)$ are found in \eqref{FGRQ}, \eqref{F} and \eqref{G}. 

Neither the energy--momentum nor the Cotton is of the perfect form, given the velocity congruence 
\eqref{uuPD}
\begin{equation}
\text{u}=\text{d}\varphi-
\chi^2\text{d}\tau +\frac{\text{d}\chi}{F}.\nonumber
 \end{equation}
The energy--momentum tensor $\text{T}$ can be decomposed as in Eqs. \eqref{Tdec}, \eqref{Tperf} in a perfect-fluid piece plus a deviation given by
\begin{eqnarray}
\Pi&=&-\frac{3\varepsilon}{2k^4}\left(\left(F+k^2\right)G\left(\text{d}\varphi-\chi^2\, \text{d}\tau\right)^2+G^2\left(\text{d}\tau+\chi^2\, \text{d}\varphi\right)^2+ 
k^4\frac{\text{d}\chi^2}{F^2} \right)\\
&&-\left(\text{d}\varphi-\chi^2\, \text{d}\tau\right)\left(3\varepsilon\frac{\text{d}\chi}{F}+\frac{\kappa c}{k^7}FG\left(\text{d}\tau+\chi^2\, \text{d}\varphi\right)\right).
\end{eqnarray}
This tensor contains all physical information regarding the boundary fluid: viscous hydrodynamic and non-hydrodynamic modes. It has been obtained here from purely boundary considerations, following integrability requirements.

\section{The reconstructed null tetrad for Pleba\'nski--Demia\'nski}\label{null}

We display here some intermediate elements necessary for resumming the boundary data provided by the metric  
\eqref{PDbry}, the congruence \eqref{uuPD}, and the energy--momentum \eqref{fullT}. The resummation is performed using \eqref{papaefgentetr}, for which we need 
to determine $\mathbf{k}, \mathbf{l}$ and $\mathbf{m}$ as in Eqs. \eqref{km} and  \eqref{l}. 
We find: 
\begin{equation}
\label{uPD}
\mathbf{k}=-\text{u}=
\chi^2\text{d}\tau-\text{d}\varphi- \frac{\text{d}\chi}{F},
\end{equation}
and 
\begin{equation}
\label{mPD}
\mathbf{m}=-\sqrt{\frac{G}{2}}\, \rho
\left(\sqrt{\chi^4+1}(\text{d}\tau+i\, \text{d}\varphi)
+\frac{k^2\left(i-\chi^2\right)\text{d}\chi}{FG \sqrt{\chi^4+1}}\right).
\end{equation}
In order to determine $\mathbf{l}$, the following is also useful (see \eqref{curlR}):
  \begin{eqnarray}
\label{RcPD}
\mathscr{R}=\frac{k^2}{\left(1+\chi^4\right)^2}
\bigg(2\chi^2(F-3G) 
-2\chi^3
\left(1+\chi^4\right)G'
-
\left(1+\chi^4\right)^2G''
\bigg);
  \end{eqnarray}
combined with \eqref{thPD} and \eqref{qPD} in \eqref{Hgen}, this leads to
 \begin{eqnarray}
 H&=&\frac{k^2}{2}\left(r^2+\frac{5\chi^2}{\left(1+\chi^4\right)^2}\right)
 -
 \frac{2\chi^2 G}{\left(1+\chi^4\right)^2}
 \left(1+r\chi\left(1+\chi^4\right)\right)\nonumber
\\ && -\left(r+\frac{\chi^3}{1+\chi^4}\right)\frac{G'}{2}
-\frac{G''}{4} -\frac{\chi c}{2k^4\left(1+\chi^4\right)\rho^2}
-\frac{3r\varepsilon}{2k\kappa \rho^2}.
\label{HPD}
  \end{eqnarray}
Further using \eqref{uuPD} and \eqref{aPD} we obtain:
   \begin{eqnarray}
   \nonumber
\gamma&:=&
-r \text{a} +
\frac{1}{2k^2}  \ast(\text{u}\wedge (\text{d} q+q\text{a}))\\
 \nonumber
&\ =&\frac{1}{\left(1+\chi^4\right)^2}
 \bigg[
 \frac{2k^2\chi^2}{F}\left(1-r\chi\left(1+\chi^4\right)\right)\text{d}\chi \\
 &&-
G\left(\left(3\chi^4-1-2r\chi\left(1+\chi^4\right)^2\right)
\text{d}\tau+\chi^2\left(\chi^4-3\right)\text{d}\varphi
\right)\bigg].
 \label{starqPD}
  \end{eqnarray}
Finally Eq. \eqref{l} provides $\mathbf{l}$:
\begin{equation}
\label{lPD}
\mathbf{l}=-\text{d}r-H \text{u}
+
\gamma.
\end{equation}
In the above expressions $F(\chi)$ and $G(\chi)$ are given in \eqref{FGRQ}, \eqref{F}  and \eqref{G}, and the energy density $\varepsilon(\chi)$ and Cotton projection $c(\chi)$ are displayed in \eqref{epsPD} and \eqref{cPD}.


\end{document}